# PART I : A SIMPLE GENERAL SOLUTION OF THE RADIAL SCHRÖDINGER EQUATION FOR SPHERICALLY SYMMETRIC POTENTIALS

Hasan Huseyin Erbil[+]
Ege University, Science Faculty, Physics Department Bornova - IZMIR 35100, TURKEY

By using a simple procedure the general solution of the time-independent radial Schrödinger equation for spherical symmetric potentials was made without making any approximation. The wave functions are always periodic. It appears to be two difficulties: one of them is the solution of the equation E = U(r), where E and U(r) are the total and effective potential energies, respectively, and the other is the calculation of the integral $\int \sqrt{U(r)}\, dr$ .If analytical calculations are not possible, one must apply numerical methods. To find the energy and the wave function of the ground state, there is no need for the calculation of this integral, it is sufficient to find the classical turning points, that is to solve the equation E = U(r).

[+] E-mail: erbil@sci.ege.edu.tr Fax: +90 232 388 1036

## I. INTRODUCTION

Although we succeed in solving the time-independent radial Schrödinger equation for some simple spherical symmetric potentials, an exact solution is not possible in complicated situations, and we must then resort to approximation methods. For the calculation of stationary states and energy eigenvalues, these include perturbation theory, the variational method and the WKB approximation. Perturbation theory is applicable if the Hamiltonian differs from an exactly solvable part by a small amount. The variational method is appropriate for the calculation of the ground state energy if one has a qualitative idea of the form of the wave function and the WKB method is applicable in the nearly classical limit. In this study we achieved a simple method for the exact general solution of the time-independent radial Schrödinger equation for spherically symmetric potential without making any approximation. We have applied this simple method to some spherical symmetric potentials in the following works.

## II. TIME-INDEPENDENT RADIAL SCHRÖDINGER EQUATION FOR SPHERICAL SYMMETRIC POTENTIALS

The time-independent Schrödinger equation in three dimensions is given as

$$\Delta\psi(\vec{r}) + \frac{2m}{\hbar^2}\left[E - V(\vec{r})\right]\psi(\vec{r}) = 0 \quad (1)$$

Where, E and V are the total and potential energies respectively, m is the mass of particle. Spherical polar coordinates, x = r sin(θ) cos(φ), y = r sin(θ) sin(φ), z = r cos(θ), are appropriate for the symmetry of the problem. The Schrödinger equations (1), expressed in these coordinates, are

$$\left[\frac{\partial^2}{\partial r^2} + \frac{2}{r}\frac{\partial}{\partial r}\right]\psi(r,\theta,\varphi) + \frac{1}{r^2}\hat{L}^2(\theta,\varphi)\psi(r,\theta,\varphi) + \frac{2m}{\hbar^2}\left[E - V(r,\theta,\varphi)\right]\psi(r,\theta,\varphi) = 0 \quad (2)$$

where $\hat{L}^2(\theta,\varphi) = \frac{\partial^2}{\partial\theta^2} + \cot g(\theta)\frac{\partial}{\partial\theta} + \frac{1}{\sin^2(\theta)}\frac{\partial^2}{\partial\varphi^2}$.

The potential energy of a particle which moves in a central, spherically symmetric field of force depends only upon the distance r between the particle and the centre of force. Thus, the potential energy should be $V(r,\theta,\varphi) = V(r)$. Solution of the Equations (2) can be found by the method of separation of variables. To apply this method, we attempt to find a solution of the form

$$\psi(r,\theta,\varphi) = R(r)Y(\theta,\varphi) \quad (3)$$

in which R(r) is independent of the angles, and $Y(\theta,\varphi)$ is independent of r. Substituting Equation (3) into equation (2) and rearranging, we obtain:

$$\frac{\partial^2 R(r)}{\partial r^2} + \frac{2}{r}\frac{\partial R(r)}{\partial r} + \left\{\frac{2m}{\hbar^2}\left[E - V(r)\right] - \frac{C}{r^2}\right\}R(r) = 0 \quad (4a)$$

$$\hat{L}^2(\theta,\varphi)Y(\theta,\varphi) + CY(\theta,\varphi) = 0 \quad (4b)$$

Where C is a constant. Equation (4b) is independent of the total energy E and of the potential energy V(r). Therefore, the angular dependence of the wave functions is determined by the property of spherical symmetry, and admissible solutions of Equation (4b) are valid for every spherically symmetric system regardless of the special form of the potential function. The solutions of the Equation (4b) can be found in any quantum mechanics text–book and the solutions are known as spherical harmonic functions, $Y_{\ell\mu}(\theta, \varphi)$. Where $C = \ell(\ell+1)$, $\ell = 0, 1, 2, 3, ...$ are positive integers and $\mu = -\ell, -\ell+1, ..., 0, ..., +\ell$. Substituting $C = \ell(\ell+1)$ value and $F(r) = r\,R(r)$ value into Equation (4a) , we obtain the radial wave equation as follows:

$$\left\{\frac{\partial^2}{\partial r^2} + \frac{2m}{\hbar^2}\left[E - U(r)\right]\right\}F(r) = 0 \tag{5}$$

Where $U(r) = V(r) + \dfrac{\hbar^2}{2m}\dfrac{\ell(\ell+1)}{r^2}$ is the effective potential energy. This effective potential energy function is shown graphically in Figure 1.

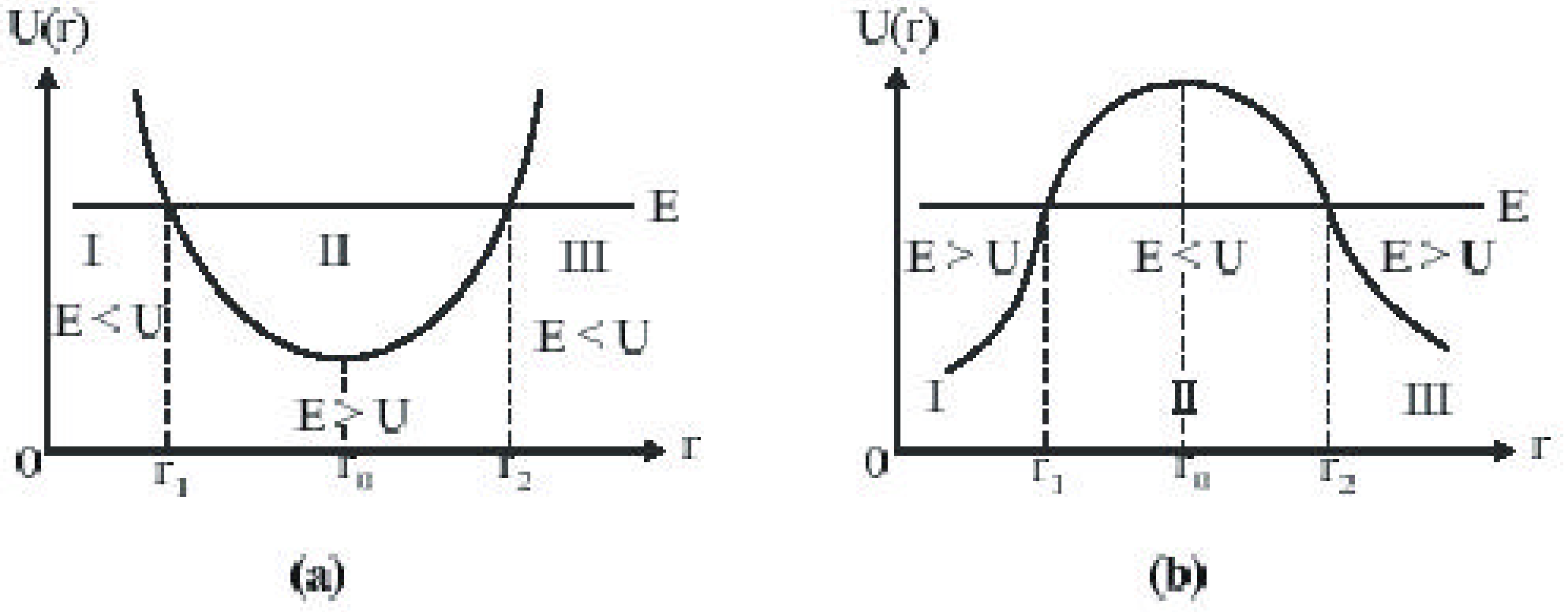

**Figure 1.** Domains relevant to a particle of energy E moving in a central, spherically symmetric field of potential U(r). **(a)** In the domains I and III; E < U(r), so kinetic energies are negative: in the domain II; E > U(r), so kinetic energies are positive: **(b)** In the domains I and III, E > U(r), so kinetic energies are positive: In the domain II, E < U(r), so kinetic energy is negative. $r_1$ and $r_2$ are the roots of the equation E = U(r) and they are turning points of the corresponding classical motion.

## III. SOLUTION OF THE RADIAL SCHRÖDINGER EQUATIONS

### III.1. The Case of E > U(r) (the kinetic energy is positive, the bound state or scattering )

#### III.1.1. Solution For The Ground State

In this case, we should solve the differential Equation (5). To solve this equation for the ground state, let us perform the following transformations:

$$U(r) \rightarrow S\delta(r - r_0) \quad \text{and} \quad F(r) \rightarrow D(r) \tag{6}$$

Where $$S = \int_{r_1}^{r_2} U(r)\, dr \tag{7}$$

Here $\delta(r-r_0)$ is Dirac function and S is the area between the graph of U (r) ) and the r axis on $[r_1, r_2]$, $r_1$ and $r_2$ are the roots of the Equation E = U(r) and $r_0 = \frac{r_1 + r_2}{2}$. If we take $d = r_2 - r_1$, then we find $r_1 = r_0 - d/2$, $r_2 = r_0 + d/2$ ( See Figure 1). With this transformation the radial Equation (5) becomes:

$$\frac{d^2 D(r)}{dr^2} + \frac{2m}{\hbar^2} E\, D(r) = +\frac{2m}{\hbar^2}\, S\, \delta\left(r - r_0\right) D(r) \tag{8}$$

To evaluate the behaviour of D(r) at r = $r_0$, let us integrate the equation (8) over the interval ($r_0-\varepsilon$, $r_0+\varepsilon$) and let us consider the limit $\varepsilon \to 0$. We obtain

$$D'(r_0+\varepsilon) - D'(r_0-\varepsilon) = \frac{2m}{\hbar^2} S\, D(r_0) \tag{9}$$

Equation (9) shows that the derivative of D(r) is not continuous at the r = $r_0$ point [1,2]. On the other hand, the wave function, F(r), should be continuous at the r = $r_0$ point.

Now, to solve the differential equation (8), we can take the Laplace's transformation of the Equation (8). LT [D(r)] = Z(q) is Laplace's Transformation of D(r). From (8):

$$q^2 LT\left[D(r)\right] + \frac{2m}{\hbar^2} E\, LT\left[D(r)\right] = +\frac{2m}{\hbar^2}\, S\, LT\left[D(r)\ \delta\left(x - x_0\right)\right]$$

$$q^2 Z(q) + \frac{2m}{\hbar^2} E\, Z(q) = \frac{2m}{\hbar^2}\, S\, e^{-qr_0} D(r_0)$$

From this equation, we get

$$Z(q) = a^2 \frac{e^{-qr_0}}{q^2 + K^2}, \quad \text{where,} \quad K^2 = \frac{2m}{\hbar^2} E \quad \text{and} \quad a^2 = \frac{2m}{\hbar^2} S\, D(r_0) \tag{10}$$

The function D(r) can be obtained by the inverse Laplace's Transformation of Z(q),

$$D(r) = LT^{-1}\left[Z(q)\right] = \frac{a^2}{2iK}\left[e^{iK(r-r_0)} - e^{-iK(r-r_0)}\right]$$

$$= A e^{iK(r-r_0)} + A^* e^{-iK(r-r_0)} \tag{11a}$$

Where, we have supposed that E > 0 and $K^2 > 0$. If we take that E < 0 and $K^2 < 0$, then K is imaginary. If we put $k = i K$, (11a) can be written as

$$D(r) = A e^{k(r-r_0)} + A^* e^{-k(r-r_0)} \tag{11b}$$

If we take $A^* = B$ in the equations (11), then we can rewrite them as follows:

$$D(r) = A e^{k(r-r_0)} + B e^{-k(r-r_0)} \tag{12}$$

A and B are integral constants to be determined by the boundary conditions. From (12) ,

$D(r) = A e^{k(r-r_0)}$ and $D(r) = A e^{-k(r-r_0)}$ are also the solutions of the Equation (8). Therefore, we get

$$D(r) = A e^{k(r-r_0)} \quad \text{for} \quad r < r_0 \tag{13a}$$

$$D(r) = A e^{-k(r-r_0)} \quad \text{for} \quad r > r_0 \tag{13b}$$

These functions could also be found by the Transformation of Fourier. Inserting the functions (13) into the equation (9) and taking the limit for $\varepsilon \to 0$, we have:

$$k = \frac{m}{\hbar^2} S \qquad \text{or} \qquad E = -\frac{m}{2\hbar^2} S^2 \tag{14}$$

To find the constant A, the function D(r) can be normalized to 1:

$$\int_{-\infty}^{r_0} AA^* e^{2k(r-r_0)}\, dr + \int_{r_0}^{+\infty} AA^* e^{-2k(r-r_0)}\, dr = 1$$

From this equation, we find the constant A as:

$$|A| = \sqrt{k} = \frac{\sqrt{mS}}{\hbar} \tag{15}$$

From (13a) and (13b), by addition and subtraction, we can get the following functions:

$$D(r) = \frac{1}{2} A\left[e^{k(r-r_0)} + e^{-k(r-r_0)}\right] = A\cosh\left[k(r-r_0)\right] \tag{16a}$$

$$D(r) = \frac{1}{2} A\left[e^{k(r-r_0)} - e^{-k(r-r_0)}\right] = A\sinh\left[k(r-r_0)\right] \tag{16b}$$

Where A is the normalization constant to be determined.

### III.1.2. Solution For The Excited States

Now, to find the general solution of the Equation (5) for the excited states, let us accept the wave function F(r) to be $F(r) = D(r)\, e^{iG(r)}$ and we put this function in Equation (5):

$$D''(r) - D(r)G'^2(r) - k^2 D(r) - m_1^2 U(r)D(r) + i\left[2D'(r)G'(r) + D(r)G''(r)\right] = 0 \tag{17}$$

Where, $m_1^2 = \frac{2m}{\hbar^2}$ and $k^2 = -\frac{2m}{\hbar^2} E$. Taking the real and imaginary parts of the equation (17), we get:

$$D''(r) - D(r)G'^2(r) - k^2 D(r) - m_1^2 U(r)D(r) = 0 \tag{18a}$$

$$2D'(r)G'(r) + D(r)G''(r) = 0 \tag{18b}$$

To find the function G(r), we can use the Equation (18). Let us take the function

$D(r) = A\cosh\left[k(r-r_0)\right]$, and its first and second derivatives $V(r) = \frac{1}{2} m\omega^2 r^2$ and

$$D''(r) = k^2 A\cosh\left[k(r-r_0)\right] = k^2 D(r)$$

Inserting these functions into (18a), we get

$$k^2 D(r) - D(r)G'^2(r) - k^2 D(r) - m_1^2 U(r)D(r) = 0$$

Since $D(r) \neq 0$, dividing by D(r), we have:

$$G'^2(r) = -m_1^2 U(r) \qquad \text{or} \tag{19a}$$

$$G'(r) = \pm\sqrt{-m_1^2 U(r)} = \pm i\, m_1 \sqrt{U(r)} \tag{19b}$$

From Equation (19b), we get

$$G(r) = \pm i\, m_1 \int \sqrt{U(r)}\, dr + C \tag{20}$$

Where, C is any constant to be determined. On the other hand, by differentiating the equation (19a), we can get G"(r):

$$2G'(r)G''(r) = -m_1^2 U'(r) \quad \text{or} \quad G''(r) = -\frac{m_1^2}{2G'(r)} U'(r)$$

We replace this value of G"(r) in to (18b)

$$2D'(r)G'(r) - D(r)\frac{m_1^2}{2G'(r)} U'(r) = 0$$

Since $D'(r) = kA \sinh[k(r - r_0)]$, we get

$$[G'(r)]^2 = \frac{m_1^2}{4k} \coth[k(r - r_0)] U'(r) \tag{21a}$$

$$G'(r) = \pm\frac{m_1}{2}\frac{1}{\sqrt{k}}\sqrt{U'(r)\coth[k(r - r_0)]} \tag{21b}$$

Integrating Equation (21b), we have

$$G(r) = \pm\frac{m_1}{2}\frac{1}{\sqrt{k}}\int \sqrt{U'(r)\coth[k(r - r_0)]}\, dr + C \tag{22}$$

Multiplying Equation (18b) by D(r), we can rewrite it

$$\frac{d}{dr}\left[D^2(r)G'(r)\right] = 0. \text{ From this, we get } D^2(r)G'(r) = \text{constant} = b \tag{23}$$

From Equation (23), we obtain

$$G'(r) = \frac{b}{D^2(r)} \tag{24}$$

Integrating Equation (24), we have:

$$G(r) = \frac{b}{kA^2} \tanh[k(r - r_0)] + C \tag{25}$$

Finally, inserting $D'(r) = kA \sinh[k(r - r_0)]$ in Equation (18b) and integrating it we get

$$G(r) = \tanh[k(r - r_0)] + C \tag{26}$$

Comparing Equation (26) with Equation (25), we find that $b = k\, A^2$. Therefore, we have three functions for G(r). We rewrite these functions as follows:

$$G(r) = \pm i\, m_1 \int \sqrt{U(r)}\, dr + C \tag{27a}$$

$$G(r) = \pm\frac{m_1}{2}\frac{1}{\sqrt{k}}\int \sqrt{U'(r)\coth[k(r - r_0)]}\, dr + C \tag{27b}$$

$$G(r) = \tanh\left[k(r - r_0)\right] + C \tag{27c}$$

Thus, for the wave function F(r) , we have

$$F(r) = D(r)\left[A e^{iG(r)} + B e^{-iG(r)}\right] \tag{28}$$

Where, A and B are the constants to be determined by the boundary conditions.

Using the same procedure, by the different D(r) functions, we can obtain the different G(r) functions which are given in Table 1.

**Table 1.** The elements of the wave functions in the case E > U(r).

| **D(r)** | **G(r)** |
|---|---|
| $A \cosh\left[k(r - r_0)\right]$ | $\pm i\, m_1 \int \sqrt{U(r)}\, dr$ |
| | $\hbar \qquad \omega$ |
| | $\frac{b}{kA^2} \tanh\left[k(r - r_0)\right]$ |
| $A \sinh\left[k(r - r_0)\right]$ | $D(r) = \frac{1}{2} A\left[e^{k(r-r_0)} - e^{-k(r-r_0)}\right] = A \sinh\left[k(r - r_0)\right]$ |
| | $\pm \frac{m_1}{2} \frac{1}{\sqrt{k}} \int \sqrt{U'(r) \tanh\left[k(r - r_0)\right]}\, dr$ |
| | $-\frac{b}{kA^2} \coth\left[k(r - r_0)\right]$ |
| $A e^{k(r - r_0)}$ | $\pm i\, m_1 \int \sqrt{U(r)}\, dr$ |
| | $\pm \frac{m_1}{2} \frac{1}{\sqrt{k}} \int \sqrt{U'(r)}\, dr$ |
| | $-\frac{b}{2k} e^{-2k(r - r_0)}$ |
| $A e^{-k(r - r_0)}$ | $\pm i\, m_1 \int \sqrt{U(r)}\, dr$ |
| | $\pm \frac{m_1}{2} \frac{i}{\sqrt{k}} \int \sqrt{U'(r)}\, dr$ |
| | $\frac{b}{2k} \frac{1}{D^2(r)}$ |
| $F(r) = D(r)\left[A e^{iG(r)} + B e^{-iG(r)}\right]$ and $F(r) = A e^{k(r - r_0) + iG(r)} + B e^{-k(r - r_0) - iG^*(r)}$ | |
| $D^2(r) G'(r) = \text{constant} = b,\ i = \sqrt{-1},\ m_1 = \sqrt{\frac{2m}{\hbar^2}},\ k = \sqrt{-\frac{2m}{\hbar^2} E} = i m_1 \sqrt{E}$ | |

| |
|---|
| If we take kr instead of $k(r-r_0)$ in both D(r) and G(r), they are also the solutions. |

### III.2. The Case E < U(r) (the kinetic energy is negative and classically not acceptable, the unbound state or tunneling)

To obtain the wave functions in this case, it is sufficient to replace $-k^2$ with $k^2$ (or ik instead of k) and to use $-m_1^2$ instead of $m_1^2$ ( or i $m_1$ instead of $m_1$) in the above functions.

## IV. BOUNDARY CONDITIONS

Let us divide the potential domain in three parts, shown as in Figure 1. In each domain, we show the wave functions as $F_1(r)$, $F_2(r)$ and $F_3(r)$. The wave functions and their derivatives should be continuous. Since the wave functions and their derivatives are continous, the above functions must obey the following conditions.

$$\begin{aligned} F_1(r_1) &= F_2(r_1) \qquad & F_1'(r_1) &= F_2'(r_1) \\ F_2(r_2) &= F_3(r_2) \qquad & F_2'(r_2) &= F_3'(r_2) \end{aligned} \tag{29}$$

The normalization of the bound state requires that the functions vanish at infinity. By these boundary conditions and normalization conditions of the wave functions, we can find the integral constants, A and B, and the energy E, in the excited bound states. In the bound ground states, we do not need the solutions of the Schrödinger equation. It is sufficient to know the classical turning points, $r_1$ and $r_2$. We shall see it in the next section.

## V. DETERMINATION OF THE GROUND STATE ENERGY IN THE BOUND STATE

The kinetic energy of the particle is $\begin{matrix} F_1(r_1)=F_2(r_1) & F_1'(r_1)=F_2'(r_1) \\ F_2(r_2)=F_3(r_2) & F_2'(r_2)=F_3'(r_2) \end{matrix}$. Integrating this equation from $r_1$ to $r_2$, we get

$$\int_{r_1}^{r_2} \frac{p^2}{2m} dr = \int_{r_1}^{r_2} [E - U(r)]\, dr = E(r_2 - r_1) - S$$

**(a)** For the positive kinetic energy, $E(r_2 - r_1) - S > 0$ (the bound state)

**(b)** For the negative kinetic energy, $E(r_2 - r_1) - S < 0$ (the unbound state).

For the bound states, inside the interval $[r_1, r_2]$ the kinetic energy is positive, outside the interval $[r_1, r_2]$ the kinetic energy is negative. The minimum point of the potential function corresponds to the ground state. Thus, at the minimum point of the potential function, we can write

$$E_0(r_2 - r_1) - S = 0$$

From this equation, we find the value of S as; $S = E_0(r_2 - r_1)$

Or, looking at the Figure 1-a, we can also write $E(r_2 - r_1) - S = \int_{r1}^{r2} T dr$ for the region II, where T is kinetic energy. At the ground state, the kinetic energy is zero, namely, T = 0. Therefore, from this equation we also find the same value $E_0(r_2 - r_1) = S$. We can replace this value of S in equation (14) and we get

$$E_0 = -\frac{m}{2\hbar^2} S^2 = -\frac{m}{2\hbar^2} E_0^2 (r_2 - r_1)^2 \quad \text{or} \quad E_0 = -\frac{2\hbar^2}{m} \frac{1}{(r_2 - r_1)^2} = -\frac{2\hbar^2}{m} \frac{1}{d^2} \tag{30}$$

$E_0$ represents the ground state energy, where the negative sign indicates that the state is bound and it can be omitted in the calculations for positive energies.

**Example:** $V(r) = -\frac{Ze^2}{r}$ (hydrogen like atom potential)

$$U(r) = V(r) + \frac{\hbar^2}{2m} \frac{\ell(\ell+1)}{r^2} = -\frac{Ze^2}{r} + \frac{\hbar^2}{2m} \frac{\ell(\ell+1)}{r^2} = -\frac{a}{r} + \frac{b}{r^2}$$

with $b = \frac{\hbar^2}{2m} \ell(\ell+1)$, $a = Z e^2$. From $-\frac{a}{r} + \frac{b}{r^2} = -|E|$, we get:

$$r_1 = \frac{a - \sqrt{a^2 - 4b|E|}}{2|E|} \quad \text{and} \quad r_2 = \frac{a + \sqrt{a^2 - 4b|E|}}{2|E|}, \quad d = r_2 - r_1 = \frac{\sqrt{a^2 - 4b|E|}}{|E|}$$

$$(r_2 - r_1)^2 = d^2 = \frac{a^2 - 4b|E|}{|E|^2}.$$ From (30), we have:

$$-|E_0| = -\frac{2\hbar^2}{m} \frac{1}{d^2} = \frac{2\hbar^2}{m} \frac{|E_0|^2}{a^2 - 4b|E_0|} \quad \text{or} \quad a^2 - 4b|E_0| = \frac{2\hbar^2}{m} |E_0|$$

$$ma^2 = 4mb|E_0| + 2\hbar^2 |E_0| = [4mb + 2\hbar^2]|E_0|$$

$$|E_0| = \frac{ma^2}{4mb + 2\hbar^2} = \frac{me^4}{2\hbar^2} \frac{Z^2}{1 + \ell(\ell+1)} \quad \text{or} \quad E_0 = -\frac{me^4}{2\hbar^2} \frac{Z^2}{[1 + \ell(\ell+1)]}$$

For the ground state of the hydrogen atom; Z = 1, m is electron mass, e is electron charge and $\ell$ = 0, we get $E_0$ = –13.6 eV. This is the ground state energy of the hydrogen atom.

## VI. CONCLUSION

Examining Table 1 in detail, we have got the main solutions of the radial Schrödinger equation for the central potential as follows:

1) $R(r) = \frac{1}{r} e^{k(r - r_0)} \left[ A e^{iG(r)} + B e^{-iG(r)} \right]$

2) $R(r) = \frac{1}{r} e^{-k(r-r_0)} \left[ A e^{iG(r)} + B e^{-iG(r)} \right]$

3) $R(r) = \frac{1}{r} \cosh\left[k(r - r_0)\right] \left[ A e^{iG(r)} + B e^{-iG(r)} \right]$

4) $R(r) = \frac{1}{r} \sinh\left[k(r - r_0)\right] \left[ A e^{iG(r)} + B e^{-iG(r)} \right]$

5) $R(r) = \frac{1}{r} \left[ A e^{k(r-r_0)+iG(r)} + B e^{-k(r-r_0)-iG^*(r)} \right]$

If one takes kr instead of k(r- $r_0$) in the above functions, they are also the solutions.

Where, (a) For E > U(r), $k = i\, m_1 \sqrt{E}$ , $G(r) = i m_1 \int \sqrt{U(r)}\, dr$

(b) For E < U(r), $k = m_1 \sqrt{E}$ , $G(r) = m_1 \int \sqrt{U(r)}\, dr$

Here $m_1 = \sqrt{\frac{2m}{\hbar^2}}$ . $r_1$ and $r_2$ are the roots of the equation E = U(r), $r_0 = \frac{r_1 + r_2}{2}$ , $d = r_2 - r_1$ , $i = \sqrt{-1}$ , $A\ e^{k(r-r_0)}$ , $r_2 = r_0 + d/2$ . The bound ground state energy is given by the formula

$$E_0 = -\frac{2\hbar^2}{m} \frac{1}{(r_2 - r_1)^2} = -\frac{2\hbar^2}{m} \frac{1}{d^2}$$

The negative sign of this formula indicates the bound state and it can be omitted in the calculations for positive energies. For the bound ground state wave function, it should be taken G(r) = 0 and $k = k_0 = i m_1 \sqrt{E_0}$ in the above functions.

The functions 3) and 4) include the functions 1) and 2), because 3) and 4) are the linear combinations of 1) and 2). The functions 5) include also the functions 3) and 4). Thus, it is possible to take only the function 5) as the solution of Schrödinger equation. A and B coefficients are determined by using the boundary and normalization conditions. Although the function 5) is sufficient by itself we nevertheless use the other functions ( 1) to 5)) in some applications.

In our procedure, there may be two difficulties: one is to solve the equation E = U(r) and the other is to find the integral of the $\sqrt{U(r)}$ , namely, to do the calculation of $\int \sqrt{U(r)}\, dr$ . If one can not do these calculations analytically, they must be performed by numerical methods. To find the energy and the wave function of the ground state, there is no need to calculate this integral, it is sufficient to find the classical turning points, namely, to solve the equation E = U(r).

**Acknowledgements**

I would like to express my sincere gratitude to my wife Özel and my daughters Iþýl and Beril for their help and patience during my study as well as my colleague Prof. N. Armaðan for his help.

# PART II : SOLUTION OF THE RADIAL SCHRÖDINGER EQUATION BY A SIMPLE METHOD FOR A PARTICLE IN AN INFINITELY HIGH CENTRAL POTENTIAL WELL OF ARBITRARY FORM AND SOME EXAMPLES

Hasan Huseyin Erbil+

Ege University, Science Faculty, Physics Department Bornova - IZMIR 35100, TURKEY

By using a simple method, the time-independent radial Schrödinger equation for a particle in an infinitely high central potential well of any form was solved without making any approximation. Two different solutions were found, namely, symmetric and antisymmetric. These functions are damped periodic functions. The wave functions and corresponding energies are similar to those of the three dimensional isotropic harmonic oscillator.

+ E-mail: erbil@sci.ege.edu.tr Fax: +90 232 388 1036

## I. INTRODUCTION

The time-independent radial Schrödinger equation in spherical polar coordinates for a spherical symmetric potential is given as follows:

$$F(r) = Ae^{k(r-r_0)+iG(r)} + Be^{-k(r-r_0)-iG^*(r)}$$

where, E is the total energy, while $U(r) = V(r) + \frac{\hbar^2}{2m}\frac{\ell(\ell+1)}{r^2}$ is the effective central potential energy. F(r) is the radial part of the wave function $\psi(r,\theta,\vartheta)$, namely,

$$\psi(r,\theta,\vartheta) = R(r)Y_{\ell\mu}(\theta,\vartheta) = \frac{F(r)}{r}Y_{\ell\mu}(\theta,\vartheta) ,$$

$Y_{\ell\mu}(\theta, \vartheta)$ is the spherical harmonic functions. V(r) is the central potential.

In the previous study [1], we found the general solutions of this radial Schrödinger equation for a central potential and we got the following functions,

1) $F(r) = e^{k(r-r_0)}\left[A e^{iG(r)} + B e^{-iG(r)}\right]$

2) $F(r) = e^{-k(r-r_0)}\left[A e^{iG(r)} + B e^{-iG(r)}\right]$

3) $F(r) = \cosh\left[k(r-r_0)\right]\left[A e^{iG(r)} + B e^{-iG(r)}\right]$

4) $F(r) = \sinh\left[k(r-r_0)\right]\left[A e^{iG(r)} + B e^{-iG(r)}\right]$

5) $F(r) = A e^{k(r-r_0)+iG(r)} + B e^{-k(r-r_0)-iG^*(r)}$

Where, **(a)** For E > U(r), $k = im_1\sqrt{E}$, $G(r) = im_1\int\sqrt{U(r)}\,dr$

**(b)** For E < U(r), $k = m_1\sqrt{E}$, $G(r) = m_1\int\sqrt{U(r)}\,dr$

$m_1 = \sqrt{\dfrac{2m}{\hbar^2}}$, $r_0 = \dfrac{r_1 + r_2}{2}$; $r_1$ and $r_2$ are the roots of the equation E = U(r), $d = r_2 - r_1$, $r_1 = r_0 - d/2$, $r_2 = r_0 + d/2$. The bound ground state energy is given by the formula

$$E_0 = -\frac{2\hbar^2}{m}\frac{1}{(r_2 - r_1)^2} = -\frac{2\hbar^2}{m}\frac{1}{d^2}$$

The negative sign of this formula shows the bound state and it can be omitted in the calculations for positive energies. For the bound ground state wave function, it should be taken G(r) = 0 and $k = k_0 = im_1\sqrt{E_0}$ in the above functions.

If we take k r instead of $k\,(r - r_0)$ in the above functions, they are also the solutions. A and B are the integral constants to be determined by the boundary conditions.

## II. SOLUTION OF THE RADIAL EQUATION

### II.1. WAVE FUNCTIONS

We consider the potential energy of the particle, V(r), so that V(r) > 0 and $\lim_{r\to\infty} V(r) = +\infty$ With this central potential, the effective potential, U(r), is

$$U(r) = V(r) + \frac{\hbar^2 \ell(\ell+1)}{2m}\frac{1}{r^2} = V(r) + \frac{a}{r^2} \quad \text{with} \quad a = \frac{\hbar^2}{2m}\ell(\ell+1).$$

These potentials, V(r) and U(r), are seen in Figure 1.

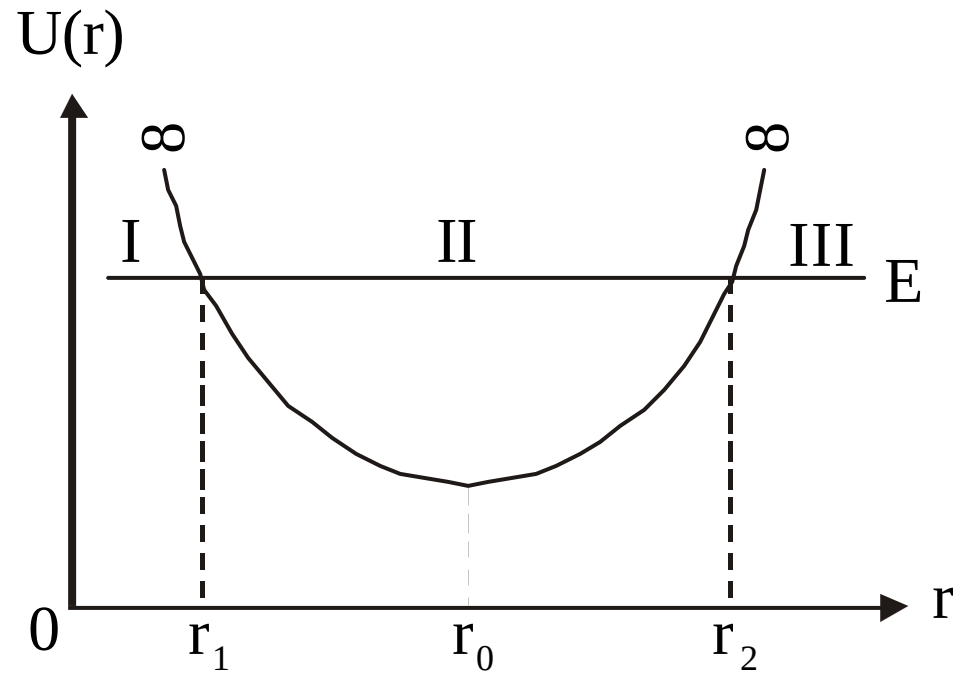

**Figure 1.** An infinitely high central effective potential well U(r). $r_1$ and $r_2$ are classical turning points.

According to Figure 1, in the domains I and III, the potentials are infinite. Therefore, the corresponding wave functions must vanish in these domains. Namely, $F_1(r) = 0$ and $F_3(r) = 0$. In the domain II, the wave function is different from zero, namely, $F_2(r) = F(r) \neq 0$. In this domain, $E > U(r)$ and $E > 0$. Thus,

$$G(r) = im_1 \int \sqrt{U(r)}\, dr = iQ(r), \qquad \text{with} \qquad Q(r) = m_1 \int \sqrt{U(r)} dr$$

We suppose that Q(r) is real, then $G^*(r) = -i\,Q(r)$. $k = im_1\sqrt{E} = iK$, with $K = m_1\sqrt{E}$

**1)** $F(r) = e^{k(r-r_0)}\left[A_1 e^{iG(r)} + B_1 e^{-iG(r)}\right]$

$$= e^{iK(r-r_0)}\left[A_1 e^{i \cdot iQ(r)} + B_1 e^{-i \cdot iQ(r)}\right] = e^{iK(r-r_0)}\left[A_1 e^{-Q(r)} + B_1 e^{Q(r)}\right]$$

In order that F(r) must vanish at $r = +\infty$; if $Q(r) > 0$, $B_1$ must be identically zero and if $Q(r) < 0$, $A_1$ must be identically zero. We suppose that $Q(r) > 0$, so $B_1 = 0$. Therefore,

$$F(r) = A_1 e^{iK(r-r_0)} e^{-Q(r)} = \{A\cos[K(r-r_0)] + B\sin[K(r-r_0)]\} e^{-Q(r)} \tag{1a}$$

**2)** $F(r) = e^{-k(r-r_0)}\left[A_1 e^{iG(r)} + B_1 e^{-iG(r)}\right]$

$$= e^{-iK(r-r_0)}\left[A_1 e^{i \cdot iQ(r)} + B_1 e^{-i \cdot iQ(r)}\right] = e^{-iK(r-r_0)}\left[A_1 e^{-Q(r)} + B_1 e^{Q(r)}\right]$$

In order that F(r) must vanish at $r = +\infty$; if $Q(r) > 0$, $B_1$ must be identically zero and if $Q(r) < 0$, $A_1$ must be identically zero. We suppose that $Q(r) > 0$, so $B_1 = 0$. Therefore,

$$D(r) = Ae^{k(r-r_0)} + Be^{-k(r-r_0)} \tag{1b}$$

**3)** $F(r) = \cosh[k(r-r_0)]\left[A_1 e^{iG(r)} + B_1 e^{-iG(r)}\right]$

$$= \cosh[iK(r-r_0)]\left[A_1 e^{i \cdot iQ(r)} + B_1 e^{-i \cdot iQ(r)}\right] = \cos[K(r-r_0)]\left[A_1 e^{-Q(r)} + B_1 e^{Q(r)}\right]$$

In order that F(r) must vanish at $r = +\infty$; if $Q(r) > 0$, $B_1$ must be identically zero and if $Q(r) < 0$, $A_1$ must be identically zero. We suppose that $Q(r) > 0$, so $B_1 = 0$. Therefore,

$$F(r) = A\cos[K(r-r_0)] e^{-Q(r)} \tag{1c}$$

**4)** $F(r) = \sinh[k(r-r_0)]\left[A_1 e^{iG(r)} + B_1 e^{-iG(r)}\right]$

$$= \sinh[iK(r-r_0)]\left[A_1 e^{i \cdot iQ(r)} + B_1 e^{-i \cdot iQ(r)}\right] = i\sin[K(r-r_0)]\left[A_1 e^{-Q(r)} + B_1 e^{Q(r)}\right]$$

In order that F(r) must vanish at $r = +\infty$; if $Q(r) > 0$, $B_1$ must be identically zero and if $Q(r) < 0$, $A_1$ must be identically zero. We suppose that $Q(r) > 0$, so $B_1 = 0$. Therefore,

$$F(r) = B\sin\left[K(r-r_0)\right]e^{-Q(r)} \tag{1d}$$

**5)** $F(r) = A_1 e^{k(r-r_0)+iG(r)} + B_1 e^{-k(r-r_0)-iG^*(r)}$

$$= A_1 e^{iK(r-r_0)+i\cdot iQ(r)} + B_1 e^{-iK(r-r_0)-i\cdot(-i)Q(r)} = \left[A_1 e^{iK(r-r_0)} + B_1 e^{-iK(r-r_0)}\right]e^{-Q(r)}$$

$$= \left\{A\cos\left[K(r-r_0)\right] + B\sin\left[K(r-r_0)\right]\right\}e^{-Q(r)} \tag{1e}$$

The functions (1a), (1b) and (1e) are the same. They are also linear combinations of (1c) and (1d). Thus, we can consider one of them and apply the boundary conditions. Boundary conditions are $F(r_1) = 0$ , $F(r_2) = 0$. From these conditions, we have

$$F(r) = \{A\cos[K(r-r_0)] + B\sin[K(r-r_0)]\}e^{-Q(r)}$$

$$F(r_1) = \{A\cos[K(r_1-r_0)] + B\sin[K(r_1-r_0)]\}e^{-Q(r_1)} = 0$$

$$F(r_2) = \{A\cos[K(r_2-r_0)] + B\sin[K(r_2-r_0)]\}e^{-Q(r_2)} = 0$$

Since $r_1 = r_0 - d/2$ , $r_2 = r_0 + d/2$ , $r_0$ , thus,

$$A\cos(K\,d/2) - B\sin(K\,d/2) = 0 \tag{2a}$$

$$A\cos(K\,d/2) + B\sin(K\,d/2) = 0 \tag{2b}$$

In order that this system of equations has a solution different from zero, the determinant of coefficients should vanish, namely,

$$\det\begin{vmatrix}\cos(K\,d/2) & -\sin(K\,d/2)\\ \cos(K\,d/2) & \sin(K\,d/2)\end{vmatrix} = 0 \quad \text{or} \quad \cos(K\,d/2)\cdot\sin(K\,d/2) = 0 \tag{3}$$

From the equation (3) we have:

**(a)** $\cos(K\,d/2) = 0$ and $\sin(K\,d/2) \neq 0$

$$\frac{K\,d}{2} = (2n-1)\frac{\pi}{2} \quad \text{or} \quad K\,d = (2n-1)\pi, \quad n = 1, 2, 3, \ldots \tag{4a}$$

Inserting this value of Kd into (2), we obtain B ≡ 0. Thus, the function should be as

$$F(r) = A\cos[K(r - r_0)]e^{-Q(r)} = A\cos\left[\frac{(2n-1)\pi}{d}(r - r_0)\right]e^{-Q(r)} \qquad (4b)$$

This is a symmetric function.

**(b)** $\cos(Kd/2) \neq 0$ and $\sin(Kd/2) = 0$

$$\frac{Kd}{2} = n\pi \quad \text{or} \quad Kd = 2\pi n, \quad n = 1, 2, 3, \ldots \qquad (5a)$$

Inserting this value of Kd into (2), we obtain A ≡ 0. Thus, the function should be as

$$F(r) = B\sin[K(r - r_0)]e^{-Q(r)} = B\sin\left[\frac{2n\pi}{d}(r - r_0)\right]e^{-Q(r)} \text{, where n = 1, 2, 3, ...} \qquad (5b)$$

This is an antisymmetric function.

**(c)** From the equation (3), we obtain

$$\sin(Kd) = 0 \quad \text{or} \quad Kd = n\pi, \quad n = 1, 2, 3, \ldots \qquad (6a)$$

Inserting this value of Kd into (2), we obtain A = 0, B ≠ 0 for even integer n values; and A ≠ 0, B = 0 for odd integer n values. Thus, we have the wave functions as follows;

$$F(r) = A\cos\left[\frac{n\pi}{d}(r - r_0)\right]e^{-Q(r)} \quad \text{for odd integer values of n} \qquad (6b)$$

$$F(r) = B\sin\left[\frac{n\pi}{d}(r - r_0)\right]e^{-Q(r)} \quad \text{for even integer values of n} \qquad (6c)$$

Since odd and even integers can be written as (2n-1) and (2n), respectively. We again obtain the wave functions as

$$F(r) = A\cos\left[\frac{(2n-1)\pi}{d}(r - r_0)\right]e^{-Q(r)} \text{ , } \quad n = 1, 2, 3, \ldots \qquad (4b')$$

$$F(r) = B\sin\left[\frac{2np}{d}(r - r_0)\right]e^{-Q(r)} \qquad , \qquad n = 1\,,\,2\,,\,3\,,\ \ldots \qquad (5b')$$

The function (4b) is the same as (1c) and the function (5b) is the same as (1d). Therefore, we have in fact two functions. One of them is symmetric, the other is antisymmetric and they are given by (4b) and (5b), respectively. We can find the coefficients A and B by the normalisation of the wave functions. Thus, the radial wave functions can be obtained as follows:

$$R(r) = \frac{F(r)}{r} = \frac{A}{r}\cos\left[\frac{(2n-1)\pi}{d}r\right]e^{-Q(r)}, \quad n = 1, 2, 3, \ldots \text{ (symmetric function)} \qquad (7a)$$

$$R(r) = \frac{F(r)}{r} = \frac{B}{r}\sin\left[\frac{2n\pi}{d}r\right]e^{-Q(r)}, \quad n = 1, 2, 3, \ldots \text{ (antisymmetric function)} \qquad (7b)$$

## II.2. ENERGY VALUES

### II.2.1. Ground State Energy

To determine ground state energy (or zero point energy) we can use the formula

$$E = \frac{2\hbar^2}{m}\frac{1}{(r_2 - r_1)^2} = \frac{2\hbar^2}{m}\frac{1}{d^2} \qquad (8)$$

### II.2.2. Excited State Energies

To determine excited state energies, we can use the equations (4a), (5a) and (6a)

**(a)** From (4a), we can have the symmetric state energies

$$K^2 d^2 = (2n-1)^2 \pi^2 , \qquad \frac{2m}{\hbar^2} E_n d^2 = (2n-1)^2 \pi^2$$

$$E_n = \frac{2\hbar^2 \pi^2}{md^2}\left(n - \tfrac{1}{2}\right)^2 \qquad , \qquad n = 1, 2, 3, ... \qquad (9a)$$

**(b)** From (5a), we can have the antisymmetric state energies

$$K^2 d^2 = 4\pi^2 n^2 , \qquad \frac{2m}{\hbar^2} E_n d^2 = 4\pi^2 n^2$$

$$E_n = \frac{2\hbar^2 \pi^2}{md^2} n^2 \qquad , \qquad n = 1 , 2 , 3 , ... \qquad (9b)$$

**(c)** From (6a), we can have the general case energies

$$K^2 d^2 = \pi^2 n^2 , \qquad \frac{2m}{\hbar^2} E_n d^2 = \pi^2 n^2$$

$$E_n = \frac{\hbar^2 \pi^2}{2md^2} n^2 \qquad , \qquad n = 1 , 2 , 3 , ... \qquad (9c)$$

From (9c), we obtain (9a) for odd values of n and (9b) for even values of n. Thus, the energy values (9c) include both (9a) and (9b).

## III. EXAMPLES

### Example 1. Infinitely High Spherical Symmetric Square Well

We consider a particle of mass m captured in a box limited by $0 \le r \le L$. The corresponding potential can be given by

$V(r) = 0 \qquad$ for $\qquad 0 < r < L$

$V(r) = \infty \qquad$ for $\qquad r < 0$ and $r > L$

with this potential, the effective potential U(r) in the well,

$$U(r) = V(r) + \frac{\hbar^2 \ell(\ell+1)}{2m}\frac{1}{r^2} = \frac{a}{r^2} \qquad \text{with} \quad a = \frac{\hbar^2}{2m}\ell(\ell+1).$$

These potentials are seen in Figure 2.

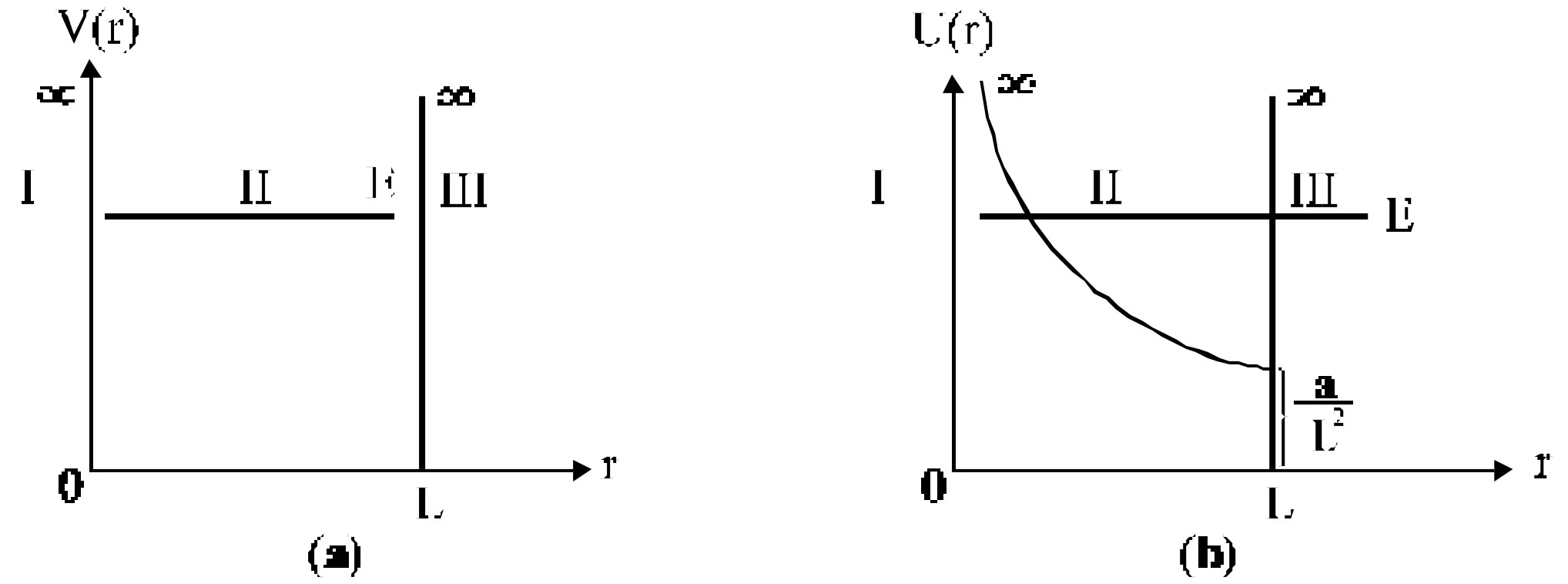

**Figure 2. (a)** The V(r) central potential **(b)** The U(r) effective central potential.

Outside the potential well, the wave functions must vanish, because U(r) is infinitely large. So, in the domains I and III, the wave functions $F_1(r) = 0$ and $F_3(r) = 0$. In the domains II, the wave function is $F_2(r) = F(r) \neq 0$. In this domain II,

$$U(r) = \frac{a}{r^2}, \quad E > 0, \quad k = im_1\sqrt{E} = i\,K, \quad \text{with} \quad K = m_1\sqrt{E}$$

$$r_1 = \sqrt{a/E} \quad , \quad r_2 = L \quad , \quad r_0 = \frac{r_2 + r_1}{2} = \frac{1}{2}\left[L + \sqrt{a/E}\right], \quad d = r_2 - r_1 = L - \sqrt{a/E}$$

$$r_1 = r_0 - d/2\,, \quad r_2 = r_0 + d/2$$

$$G(r) = im_1 \int \sqrt{U(r)}\, dr = im_1 \int \sqrt{\frac{a}{r^2}} dr = i\sqrt{\frac{2m}{\hbar^2}} a \ln(r) = iQ(r)$$

with $Q(r) = \sqrt{\frac{2m}{\hbar^2}} a \ln(r) = \sqrt{\ell(\ell+1)} \ln(r) > 0$

**Ex-1.1. Wave Functions**

From the Equations (7a) and (7b), we obtain:

$$R(r) = \frac{F(r)}{r} = \frac{A}{r} \cos\left[\frac{(2n-1)\pi}{d} r\right] e^{-Q(r)}, \qquad n = 1, 2, 3, \ldots \text{ (symmetric solution)} \qquad \text{(E1-1a)}$$

$$R(r) = \frac{F(r)}{r} = \frac{B}{r} \sin\left[\frac{2n\pi}{d} r\right] e^{-Q(r)}, \qquad n = 1, 2, 3, \ldots \text{(antisymmetric solution)} \qquad \text{(E1-1b)}$$

The known radial functions are $R_\ell(r) = A_\ell j_\ell(Kr)$; where j , spherical Bessel function [2].

**Ex-1.2. Energy Values**

**1) Ground State Energy**

To determine ground state energy (or zero point energy) we can use the Equation (8) and we obtain:

$$E_0^{(1)} = \left[\frac{\sqrt{a} + \sqrt{g}}{L}\right]^2, \; E_0^{(2)} = \left[\frac{\sqrt{a} - \sqrt{g}}{L}\right]^2 \quad \text{with} \quad g = \frac{2\hbar^2}{m} \qquad \text{(E1-2)}$$

**2) Excited State Energies**

To determine excited state energies, we can use the equations (9):

**(a)** From (9a), we can have the symmetric state energies

$$E_n^{(1)} = \left[\frac{\sqrt{a}+\sqrt{g}}{L}\right]^2, \quad E_n^{(2)} = \left[\frac{\sqrt{a}-\sqrt{g}}{L}\right]^2 \quad \text{with} \quad g = \frac{2\hbar^2\pi^2}{m}\left(n-\tfrac{1}{2}\right)^2, \; n = 1, 2, 3, \ldots \quad \text{(E1-3a)}$$

**(b)** From (9b), we can have the antisymmetric state energies

$$E_n^{(1)} = \left[\frac{\sqrt{a}+\sqrt{g}}{L}\right]^2, \quad E_n^{(2)} = \left[\frac{\sqrt{a}-\sqrt{g}}{L}\right]^2 \quad \text{with} \quad g = \frac{2\hbar^2\pi^2}{m}n^2, \quad n = 1, 2, 3, \ldots \quad \text{(E1-3b)}$$

**(c)** From (9c), we can have the general case energies

$$E_n^{(1)} = \left[\frac{\sqrt{a}+\sqrt{g}}{L}\right]^2, \quad E_n^{(2)} = \left[\frac{\sqrt{a}-\sqrt{g}}{L}\right]^2 \quad \text{with} \quad g = \frac{\hbar^2\pi^2}{2m}n^2, \quad n = 1, 2, 3, \ldots \quad \text{(E1-3c)}$$

Inserting the values of a and g into (E1-3c), we have:

$$E_{n\ell}^{(1)} = \frac{\hbar^2}{2mL^2}\left[\sqrt{\ell(\ell+1)}+\pi n\right]^2 \text{ and } E_{n\ell}^{(2)} = \frac{\hbar^2}{2mL^2}\left[\sqrt{\ell(\ell+1)}-\pi n\right]^2 \quad \text{(E1-4)}$$

The known allowed energies are given as follows [2]:

$$E_{n\ell} = \frac{\hbar^2}{2mL^2}\beta_{n\ell}^{\;2} \quad \text{(E1-5)}$$

Where $\beta_{n\ell}$, $n^{th}$ zero of the $\ell^{th}$ spherical Bessel function. Some values of $\beta_{n\ell}$ and energies calculated according to (E1-4) and (E1-5) are given in the Table 1 (unit $\frac{\hbar^2}{2mL^2}$):

**Table 1.** Some energy values of the infinitely high spherical symmetric square well

| $n\,\ell$ | $\beta_{n\ell}$ | $E_{n\ell}$ according to (E1-5) | $E_{n\ell}^{(1)}$ according to (E1-4) | $E_{n\ell}^{(2)}$ according to (E1-4) |
|---|---|---|---|---|
| 1 s | 3.142 | 9.872 | 9.870 | 9.870 |
| 1 p | 4.493 | 20.187 | 20.755 | 2.984 |
| 1 d | 5.763 | 33.212 | 31.260 | 0.479 |
| 2 s | 6.283 | 39.476 | 39.478 | 39.478 |
| 2 p | 7.725 | 59.676 | 59.250 | 23.707 |
| 2 d | 9.095 | 82.719 | 76.260 | 14.697 |

**Example 2: Three-dimensional Isotropic Harmonic Oscillator Potential**

The potential energy of three dimensional isotropic harmonic oscillator is given by $V(r)=\frac{1}{2}m\omega^2r^2$. With this potential, the effective potential U(r) is

$$U(r)=V(r)+\frac{\hbar^2\ell(\ell+1)}{2m}\frac{1}{r^2}=\frac{1}{2}mw^2r^2+\frac{\hbar^2\ell(\ell+1)}{2m}\frac{1}{r^2}=ar^2+\frac{b}{r^2}$$

with $a=\frac{1}{2}m\omega^2$, $b=\frac{\hbar^2}{2m}\ell(\ell+1)$ .These potentials, V(r) and U(r), are seen in Figure 3.

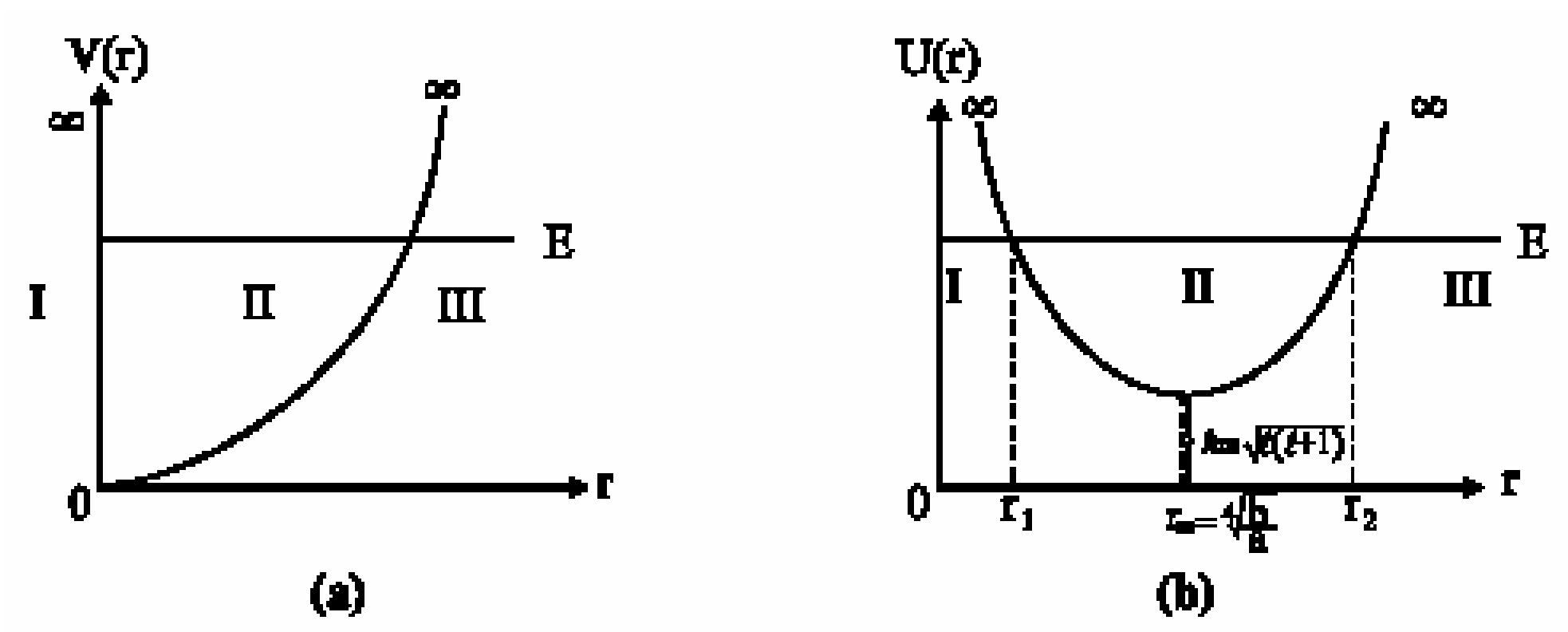

**Figure 3.** Three dimensional isotropic harmonic oscillator potential **(a)** The only harmonic oscillator potential **(b)** The effective harmonic oscillator potential

Outside the potential well, the wave functions must vanish, because U(r) is infinitely large. So, in the domains I and III, the wave functions $F_1(r) = 0$ and $F_3(r) = 0$. In the domain II, the wave function $F_2(r) = F(r) \neq 0$. In the domain II,

$$E > U(r) > 0, \quad k = im_1\sqrt{E} = iK, \quad \text{with } K = m_1\sqrt{E}$$

The positive roots of the equation E = U(r) are

$$r_1 = \sqrt{\frac{E - C}{2a}}, \quad r_2 = \sqrt{\frac{E + C}{2a}}, \quad \text{with} \quad C = \sqrt{E^2 - 4ab}$$

$$d^2 = (r_2 - r_1)^2 = \frac{E}{a} - 2\sqrt{\frac{b}{a}}, \quad r_0 = \frac{r_2 - r_1}{2}$$

$$G(r) = im_1\int\sqrt{ar^2 + \frac{b}{r^2}}\,dr = i\sqrt{\frac{2m}{\hbar^2}}\int\sqrt{ar^2 + \frac{b}{r^2}}dr$$

$$G(r) = i\,Q(r), \quad \text{with} \quad Q(r) = \sqrt{\frac{2m}{\hbar^2}}\int\sqrt{ar^2 + \frac{b}{r^2}}\,dr$$

$$Q(r)=\sqrt{\frac{2m}{\hbar^2}}\frac{1}{2}\left\{\sqrt{ar^4+b}-\sqrt{b}\ln\left[2\frac{\sqrt{b}+\sqrt{ar^4+b}}{r^2}\right]\right\}$$ is real and $G^*(r) = -i\,Q(r)$

**Ex-2.1. Wave Functions**

From the Equations (7a) and (7b), we obtain the wave functions as:

$$R(r)=\frac{F(r)}{r}=\frac{A}{r}\cos\left[\frac{(2n-1)\pi}{d}r\right]e^{-Q(r)} \quad , \quad n = 1 , 2 , 3 , ... \qquad \text{(E2-1a)}$$

$$R(r)=\frac{F(r)}{r}=\frac{B}{r}\sin\left[\frac{2n\pi}{d}r\right]e^{-Q(r)} \quad , \quad n = 1, 2, 3, ... \qquad \text{(E2-1b)}$$

The known radial wave functions are as follows:

$R_{n\ell}(\rho)=A\rho^{\ell+1}\exp(-\tfrac{1}{2}\rho^2)L_n^{\ell+1/2}(\rho^2)$. Where L is the Laguerre polinomial. (E2-1c)

***Ex-2.2. Energy Values***

**1) Ground State Energy**

To determine ground state energy (or zero point energy), we can use the Equation (8) and we obtain:

$$E_0=\frac{1}{2}\hbar\omega\left[\sqrt{\ell(\ell+1)}\pm\sqrt{\ell(\ell+1)+4}\right]$$

Since $E_0 > 0$ and the second term is greater than the first term, we should take only the (+) sing. Thus, the ground state energy is

$$E_0=\frac{1}{2}\hbar\omega\left[\sqrt{\ell(\ell+1)}+\sqrt{\ell(\ell+1)+4}\right] \qquad \text{(E2-2)}$$

**2) Excited State Energies**

To determine excited state energies, we can use the equations (9a), (9b) and (9c)

**(a)** From (9a), we can have the symmetric state energies

$$E_n = \frac{1}{2}\hbar\omega\left[\sqrt{\ell(\ell+1)} \pm \sqrt{\ell(\ell+1)+4\left(n-\tfrac{1}{2}\right)^2\pi^2}\,\right]$$

Since E > 0 and the second term is greater than the first term, we should take only the (+) sing. Thus, the symmetric state energies are

$$E_{n\ell} = \frac{1}{2}\hbar\omega\left[\sqrt{\ell(\ell+1)} + \sqrt{\ell(\ell+1)+4\left(n-\tfrac{1}{2}\right)^2\pi^2}\,\right] \;,\; n = 1\,,2\,,3\,,... \qquad \text{(E2-3a)}$$

**(b)** From (9b), we can have the antisymmetric state energies

$$E_n = \frac{1}{2}\hbar\omega\left[\sqrt{\ell(\ell+1)} \pm \sqrt{\ell(\ell+1)+4n^2\pi^2}\,\right]$$

Since $E_n > 0$ and the second term is greater than the first term, we should take only the (+) sign.

Thus, the antisymmetric state energies are

$$E_{n\ell} = \frac{1}{2}\hbar\omega\left[\sqrt{\ell(\ell+1)} + \sqrt{\ell(\ell+1)+4n^2\pi^2}\,\right] \;,\qquad n = 1\,,2\,,3\,,... \qquad \text{(E2-3b)}$$

**(c)** From (9c), we can have the general case energies

$$E_n = \frac{1}{2}\hbar\omega\left[\sqrt{\ell(\ell+1)} \pm \sqrt{\ell(\ell+1)+n^2\pi^2}\,\right]$$

Since $E_n > 0$ and the second term is greater than the first term, we should take only the (+) sign.

Thus, the general case energies are

$$E_{n\ell} = \frac{1}{2}\hbar\omega\left[\sqrt{\ell(\ell+1)} + \sqrt{\ell(\ell+1)+n^2\pi^2}\,\right] \;,\qquad n = 1\,,2\,,3\,,... \qquad \text{(E2-3c)}$$

From (E2-3c), for odd values of n, we obtain (E2-3a) and for even values of n, we get (E2-3b). Thus, the energy values (E2-3c) include both (E2-3a) and (E2-3b). It is possible to combine the equations ( E2-2 ), (E2-3a ), (E2-3b) and (E2-3c) in one equation by using $g_n$ in the following form:

$$E_{n\ell} = \frac{1}{2}\hbar\omega\left[\sqrt{\ell(\ell+1)} + \sqrt{\ell(\ell+1)+g_n}\right] \qquad \text{(E2-4)}$$

From (E2-4) we get (E2-2) for $g_n$=4 ; (E2-3a) for $g_n = 4(n-\frac{1}{2})^2 \pi^2$ ; (E2-3b) for $g_n = 4n^2\pi^2$; (E2-3c) for $g_n = n^2\pi^2$. Where n = 1 , 2 , 3 , ...

The known energy values are given as follows:

$$E_{n\ell} = (2n+\ell+3/2)\hbar\omega \quad (\text{ n}=0,1,2,3,..) \qquad \text{(E2-5)}$$

Some values of the energies , calculated according to (E2-4) and (E2-5), are given in the Table 2. ( unit ( $\hbar\omega$ )):

Table 2. Some energy values of the isotropic harmonic oscillator

| N $\ell$ | $E_{n\ell}$ according to (E2-5) | $E_{n\ell}$ according to (E2-4) |
|---|---|---|
| 1 s | 3.5 | 1.571 |
| 1 p | 4.5 | 2.430 |
| 1 d | 5.5 | 3.217 |
| 2 s | 5.5 | 3.142 |
| 2 p | 6.5 | 3.927 |
| 2 d | 7.5 | 4.597 |

**Example 3. Three-dimensional Isotropic Harmonic Oscillator With Spin-orbit Coupling**

The potential energy of a three dimensional isotropic harmonic oscillator is given by

$$V_0(r) = \frac{1}{2} m\omega^2 r^2$$

We must add the spin–orbit interaction potential to this simple potential. The spin–orbit interaction potential is given by

$$V_{s\ell}(r) = -\frac{\hbar^2}{2m^2c^2}\frac{1}{r}\frac{dV_0}{dr}\vec{\ell}\cdot\vec{s}, \quad \text{with} \quad \vec{\ell}\cdot\vec{s} = \frac{1}{2}\left[j(j+1) - \ell(\ell+1) - s(s+1)\right]$$

Where $\ell$, s and j are the orbital, spin and total angular momentum quantum numbers of a particle,

respectively. It is possible to find the derivation of this expression in any quantum mechanics text book. If we calculate $V_{s\ell}(r)$, we find the following value:

$$V_{s\ell j}(r) = -\frac{\hbar^2\omega^2}{2mc^2}\left[j(j+1) - \ell(\ell+1) - s(s+1)\right] = -C_{\ell sj} = constant$$

If we accept that $s = \frac{1}{2}$ (for fermions), we have

$$V_{s\ell j}(r) = -\frac{\hbar^2\omega^2}{2mc^2}\left[j(j+1) - \ell(\ell+1) - \tfrac{3}{4}\right] = -C_{\ell sj}$$

Thus, the harmonic oscillator potential with spin-orbit coupling becomes

$$V(r) = V_0(r) + V_{\ell sj}(r).$$

With this potential, the effective potential U(r) is

$$U(r) = V(r) + \frac{\hbar^2\ell(\ell+1)}{2m}\frac{1}{r^2} = \frac{1}{2}m\omega^2 r^2 - C_{\ell sj} + \frac{b}{r^2} = ar^2 - C_{\ell sj} + \frac{b}{r^2}$$

with $a = \frac{1}{2}m\omega^2$, $b = \frac{\hbar^2}{2m}\ell(\ell+1)$. These potentials are shown in Figure 4

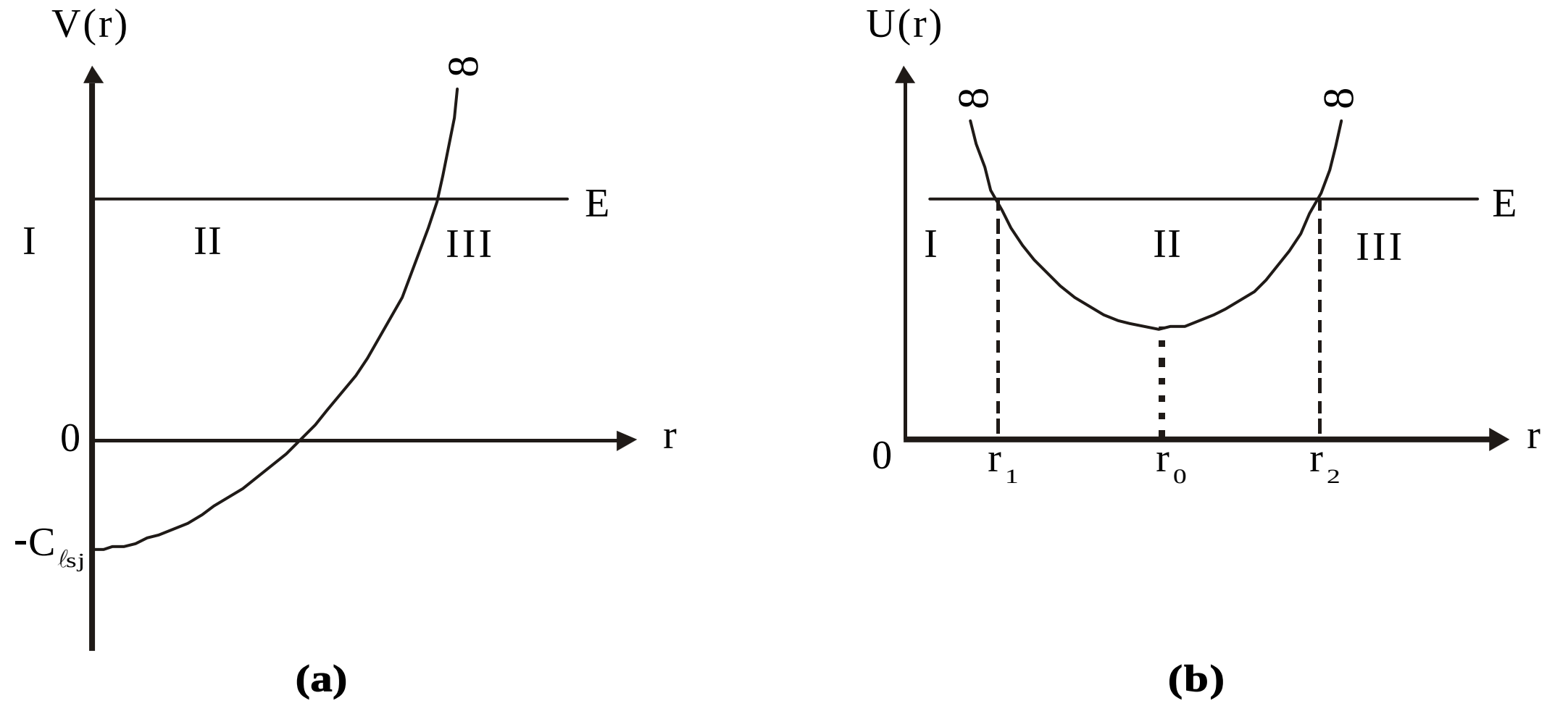

**Figure 4.** Three dimensional isotropic harmonic oscillator potential **(a)** harmonic oscillator potential with the spin–orbit coupling **(b)** The effective potential.

We consider three domains as seen in Figure 4. The wave functions for these three regions are $F_1(r)$, $F_2(r)$ and $F_3(r)$. Outside the potential well, the wave functions must vanish, because U(r) is infinitely large. Thus, in the domains I and III, the wave functions, $F_1(r) = 0$ and $F_3(r) = 0$. In the domain II, the wave function $F_2(r) = F(r) \neq 0$. Now, we will find this function.

In the domain II, $E > U(r) > 0$, $k = im_1\sqrt{E} = i\,K$, with $K = m_1\sqrt{E}$

The positive roots of the equation E = U(r) are

$$r_1 = \sqrt{\frac{(E + C_{\ell sj}) - \delta}{2a}}\,, \quad r_2 = \sqrt{\frac{(E + C_{\ell sj}) + \delta}{2a}}\,, \quad \text{with} \quad \delta = \sqrt{(E + C_{\ell sj})^2 - 4ab}$$

$$d^2 = (r_2 - r_1)^2 = \frac{(E + C_{\ell sj}) - \sqrt{(E + C_{\ell sj})^2 - \delta^2}}{a} = \frac{(E + C_{\ell sj}) - 2\sqrt{ab}}{a} \quad , \quad r_0 = \frac{r_1 + r_2}{2}$$

$$G(r) = im_1 \int \sqrt{U(r)}dr = i\sqrt{\frac{2m}{\hbar^2}} \int \sqrt{ar^2 + \frac{b}{r^2} - C_{\ell sj}}\; dr = iQ(r)$$

with $Q(r) = \sqrt{\frac{2m}{\hbar^2}} \int \sqrt{ar^2 + \frac{b}{r^2} - C_{\ell sj}}\; dr$ . Or after integrating, we have:

$$Q(r) = \sqrt{\frac{2m}{\hbar^2}} \left\{ \frac{1}{2}\sqrt{ar^4 - C_{\ell sj} r^2 + b} - \frac{C_{\ell sj}}{4\sqrt{b}} \ln\left[2\sqrt{b\left(ar^4 - C_{\ell sj} r^2 + b\right)} + 2br^2 - C_{\ell sj}\right] \right.$$
$$\left. - \frac{\sqrt{a}}{2} \ln\left[\frac{2\sqrt{a\left(ar^4 - C_{\ell sj} r^2 + b\right)} - C_{\ell sj} r^2 + 2a}{r^2}\right] \right\}$$

**Ex-3.1. Wave Functions**

From (7a) and (7b), we obtain the wave functions as:

$$\mathrm{R}(\mathrm{r}) = \frac{\mathrm{F}(\mathrm{r})}{\mathrm{r}} = \frac{\mathrm{A}}{\mathrm{r}} \cos\left[\mathrm{K}(\mathrm{r} - \mathrm{r}_0)\right] \mathrm{e}^{-\mathrm{Q}(\mathrm{r})}, \quad \text{n=1,2,3,..} \quad \text{( symmetric solution)} \qquad \text{(E3-1a)}$$

$$\mathrm{R}(\mathrm{r}) = \frac{\mathrm{F}(\mathrm{r})}{\mathrm{r}} = \frac{\mathrm{B}}{\mathrm{r}} \sin\left[\mathrm{K}(\mathrm{r} - \mathrm{r}_0)\right] \mathrm{e}^{-\mathrm{Q}(\mathrm{r})}, \quad \text{n=1,2,3,..} \quad \text{(antisymmetric solution)} \qquad \text{(E3-1b)}$$

The known radial wave functions are as follows [3]:

$$R_{n\ell}(\rho) = A\rho^{\ell+1} \exp(-\tfrac{1}{2}\rho^2) L_n^{\ell+1/2}(\rho^2).$$

Here, A, normalization constant; L, Laguerre polynomials and $\rho = \sqrt{\frac{m\omega}{\hbar}} r$ .

**Ex-3.2. Energy Values**

**1) Ground State Energy**

To determine ground state energy (or zero point energy) we can use the Equation (8) and we obtain:

$$E = \frac{1}{2}\left[ 2\sqrt{ab} - C_{\ell sj} \pm \sqrt{(2\sqrt{ab} - C_{\ell sj})^2 + \frac{8\hbar^2 a}{m}} \right]$$

Since E>0 and the term after (±) is greater than the term before it in this Equation , we should take only the (+) sign. Thus, the ground state energy is,

$$E_0 = \frac{1}{2}\left[ \sqrt{\ell(\ell+1)}\hbar w - C_{\ell sj} + \sqrt{(\sqrt{\ell(\ell+1)}\hbar w - C_{\ell sj})^2 + 4\hbar^2 w^2} \right] \qquad \text{( E3-2)}$$

**2) Excited State Energies**

To determine the excited state energies, we can use the equations (9):

(a) From (9a), we can have the symmetric state energies as follows:

$$E_n = \frac{1}{2}\left[ 2\sqrt{ab} - C_{\ell sj} \pm \sqrt{(2\sqrt{ab} - C_{\ell sj})^2 + 4\frac{\hbar^2 p^2 a(2n-1)^2}{2m}} \right]$$

Since $E_n$>0 and the term after (±) is greater than the term before it in the last equation, we should take only (+) sign. Thus the symmetric excited state energies are:

$$\mathrm{E}_{n\ell j} = \frac{1}{2}\left[ \sqrt{\ell(\ell+1)}\hbar\omega - \mathrm{C}_{\ell sj} + \sqrt{(\sqrt{\ell(\ell+1)}\hbar\omega - \mathrm{C}_{\ell sj})^2 + \pi^2(2n-1)^2\hbar^2\omega^2} \right] \qquad \text{(E3-3a)}$$

(b) From (9b), we can have the antisymmetric state energies as follows:

$$E_n = \frac{1}{2}\left[ 2\sqrt{ab} - C_{\ell sj} \pm \sqrt{(2\sqrt{ab} - C_{\ell sj})^2 + 4g_n} \right] \text{ with } g_n = n^2ð^2$$

Since $E_n$>0 and the term after (±) is greater than the term before it in the last equation, we should take only (+) sign. Thus the antisymmetric excited state energies are:

$$E_{nl} = \frac{1}{2}\left[\sqrt{\ell(\ell+1)}\hbar w - C_{\ell sj} + \sqrt{(\sqrt{\ell(\ell+1)}\hbar w - C_{\ell sj})^2 + 4g_n}\right] \quad \text{with } g_n = n^2ð^2 \qquad \text{(E3-3b)}$$

(c) From (9c), we can have the general case energies as follows:

$$E_{nl} = \frac{1}{2}\left[\sqrt{\ell(\ell+1)}\hbar w - C_{\ell sj} + \sqrt{(\sqrt{\ell(\ell+1)}\hbar w - C_{\ell sj})^2 + 4g_n}\right] \quad \text{with } g_n = n^2ð^2/4 \qquad \text{(E3-3c)}$$

It is possible to combine the equation (E3-2), (E3-3a), (E3-3b) and (E3-3c) in one equation by using

$$C_{\ell sj} = \frac{\hbar w}{2mc^2}\left[j(j+1) - \ell(\ell+1) - s(s+1)\right]\hbar w = C_j \hbar w \qquad \text{in the following form:}$$

$$\mathrm{E}_{n\ell j} = \frac{1}{2}\hbar\omega\left[\sqrt{\ell(\ell+1)} - \mathrm{C}_j + \sqrt{(\sqrt{\ell(\ell+1)} - \mathrm{C}_j)^2 + 4\mathrm{g}_n}\right] \qquad \text{(E3-4)}$$

Where, $g_n$=1 for the ground state energy (equation (E3-2)); $g_n$= $ð^2$ $(n-1/2)^2$ for the excited symmetric state energies ( equation (E3-3a)); $g_n$= $ð^2n^2$ for the excited antisymmetric state energies (equation (E3-3b)); $g_n$= $ð^2n^2/4$ for both the excited symmetric and antisymmetric state energies (equation (E3-3c). Where n=1,2,3,…

The known energy values (with first order perturbation) are as follows [3]:

$$E_{n\ell} = (2n + \ell + 3/2)\hbar\omega - \frac{C_0}{2\hbar\omega}\left[j(j+1) - \ell(\ell+1) - s(s+1)\right]\hbar\omega \quad \text{(n=0,1,2,..)} \qquad \text{(E3-5)}$$

Where, $C_0$ is a positive parameter. Some energy values calculated according to (E3-4) and (E3-5), with the values $C_0 = 0.015\hbar w$, s=1/2 and $C_j = \frac{C_0}{2\hbar w}\left[j(j+1) - \ell(\ell+1) - s(s+1)\right]$, are in Table 3

**Table 3**. Some energy values of the isotropic harmonic oscillator with spin-orbit coupling

| States | According to (E3-5) (unit $\hbar\omega$) | According to (E3-4) (unit $\hbar\omega$) |
|---|---|---|
| $1d_{5/2}$ | 3.485 | 3.204 |
| $1f_{5/2}$ | 4.530 | 4.096 |
| $1f_{7/2}$ | 4.478 | 4.051 |
| $1g_{7/2}$ | 5.538 | 5.003 |
| $1g_{9/2}$ | 5.470 | 4.941 |

**Example 4. Arbitrary Parabolic Potential Well**

We consider the following potential V(r) = a $r^2$ + b r + c. Where a, b and c are positive constants. The corresponding effective potential, U(r) is

$$U(r) = ar^2 + br + c + \frac{\delta}{r^2}, \quad \text{with} \quad \delta = \frac{\hbar^2}{2m}\ell(\ell+1)$$

These potentials may be depicted like to the Figure 4.

This potential is infinitely large and positive. Thus, in the domain II

$$G(r) = im_1 \int \sqrt{U(r)}\, dr = iQ(r), \quad \text{with} \quad Q(r) = m_1 \int \sqrt{ar^2 + br + c + \frac{d}{r^2}}\, dr$$

From the equation E = U(r), we obtain

$$ar^4 + br^3 + (c - E)r^2 + \delta = 0$$

We must accept two of the four roots of this equation. Let two roots be $r_1$ and $r_2$.

Thus, $r_0 = \frac{r_1 + r_2}{2}$, $d = r_2 - r_1$

**Ex-4.1. Wave Functions**

From (7a) and (7b), we obtain the wave functions as

$$R(r) = \frac{F(r)}{r} = \frac{A}{r}\cos\left[\frac{(2n-1)\pi}{d}(r - r_0)\right]e^{-Q(r)}, \quad (\text{n}= 1,2,3,\ldots) \quad \text{symmetric function (E4-1a)}$$

$$R(r) = \frac{F(r)}{r} = \frac{B}{r}\sin\left[\frac{2n\pi}{d}(r - r_0)\right]e^{-Q(r)}, \quad (\text{n}= 1,2,3,\ldots) \quad \text{antisymmetric function(E4-1b)}$$

**Ex-4.2. Energies Values**

1. **Ground State Energy**

To determine ground state energy, we can use the Equation (8) and we obtain:

$$E_0 = \frac{2\hbar^2}{m}\frac{1}{d^2} \qquad \text{(E4-2)}$$

2. **Exited State Energies**

(a) From (9a), we can have the symmetric state energies as follows:

$$E_n = \frac{2\hbar^2\pi^2}{md^2}\left(n-\tfrac{1}{2}\right)^2, \qquad n = 1, 2, 3, ... \qquad \text{(E4-3a)}$$

(b) From (9b), we can have the antisymmetric state energies as follows:

$$E_n = \frac{2\hbar^2\pi^2}{md^2}n^2, \qquad n = 1, 2, 3, ... \qquad \text{(E4-3b)}$$

(c) From (9c), we can have the general case energies as follows:

$$E_n = \frac{\hbar^2\pi^2}{2md^2}n^2, \qquad n = 1, 2, 3, ... \qquad \text{(E4-3c)}$$

## IV. CONCLUSION

By using our procedure, we have solved the radial Schrödinger equation for a particle found in an infinitely high central potential well of any form and we have found two different solutions. One of them is symmetric function and the other is antisymmetric function. They are given by the equations (7a) and (7b), respectively. Their corresponding energy values are also given by (9a) and (9b), respectively. Ground state energy is given by the equation (8). From these expressions, we observe that these are similar damped periodic functions. It is said that all infinitely high potential wells are similar to each other in quantum mechanics textbooks, but when we look at their solutions, it is seen that all solutions are different in form. However, in my solutions, they are similar to each other in form.

## PART III : SOLUTION OF THE RADIAL SCHRÖDÝNGER EQUATION BY A SIMPLE METHOD FOR A FREE PARTICLE

Hasan Huseyin Erbil
Ege University, Science Faculty, Physics Department Bornova - IZMIR 35100, TURKEY

By using a simple method, the time-independent radial Schrödinger equation in the spherical coordinates was solved for a free particle without making any approximation. We obtained the simple wave functions that are not dependent on the spherical Bessel and Neumann functions.

## I. INTRODUCTION

The time-independent radial Schrödinger equation in spherical polar coordinates for a spherical symmetric potential is given as follows:

$$\frac{d^2F(r)}{dr^2}+\frac{2m}{\hbar^2}[E-U(r)]F(r)=0$$

where, E is the total energy, while $U(r)=V(r)+\frac{\hbar^2}{2m}\frac{\ell(\ell+1)}{r^2}$ is the effective central potential energy. F(r) is the radial part of the wave function $\psi(r,\theta,\vartheta)$, namely,

$$\psi(r,\theta,\vartheta)=R(r)Y_{\ell\mu}(\theta,\vartheta)=\frac{F(r)}{r}Y_{\ell\mu}(\theta,\vartheta),$$

$Y_{\ell\mu}(\theta,\vartheta)$ is the spherical harmonic functions. V(r) is the central potential.

In the previous study [1], we found the general solutions of this radial Schrödinger equation for a central potential and we obtained some functions. Four of them are the following functions:

1) $F(r)=e^{k(r-r_0)}\left[Ae^{iG(r)}+Be^{-iG(r)}\right]$

2) $F(r)=e^{-k(r-r_0)}\left[Ae^{iG(r)}+Be^{-iG(r)}\right]$

3) $F(r)=\cosh\left[k(r-r_0)\right]\left[Ae^{iG(r)}+Be^{-iG(r)}\right]$

4) $F(r)=\sinh\left[k(r-r_0)\right]\left[Ae^{iG(r)}+Be^{-iG(r)}\right]$

Where, **(a)** For E > U(r), $k=im_1\sqrt{E}$, $G(r)=im_1\int\sqrt{U(r)}\,dr$

**(b)** For E < U(r), $k=m_1\sqrt{E}$, $G(r)=m_1\int\sqrt{U(r)}\,dr$

with $m_1=\sqrt{\frac{2m}{\hbar^2}}$. If we take kr instead of k(r–$r_0$) in the above functions, they are also solutions. A and B are the integral constants to be determined by the boundary conditions.

## II. SOLUTION OF THE RADIAL EQUATION

### II.1. WAVE FUNCTIONS

A free particle of mass m moves in the interval [ $0 \le r \le \infty$]. Its potential energy is taken as zero, namely, V(r) = 0. Thus, the effective potential is

$U(r)=\frac{a}{r^2}$, with $a=\frac{\hbar^2}{2m}\ell(\ell+1)$. These potentials are seen in Figure1.

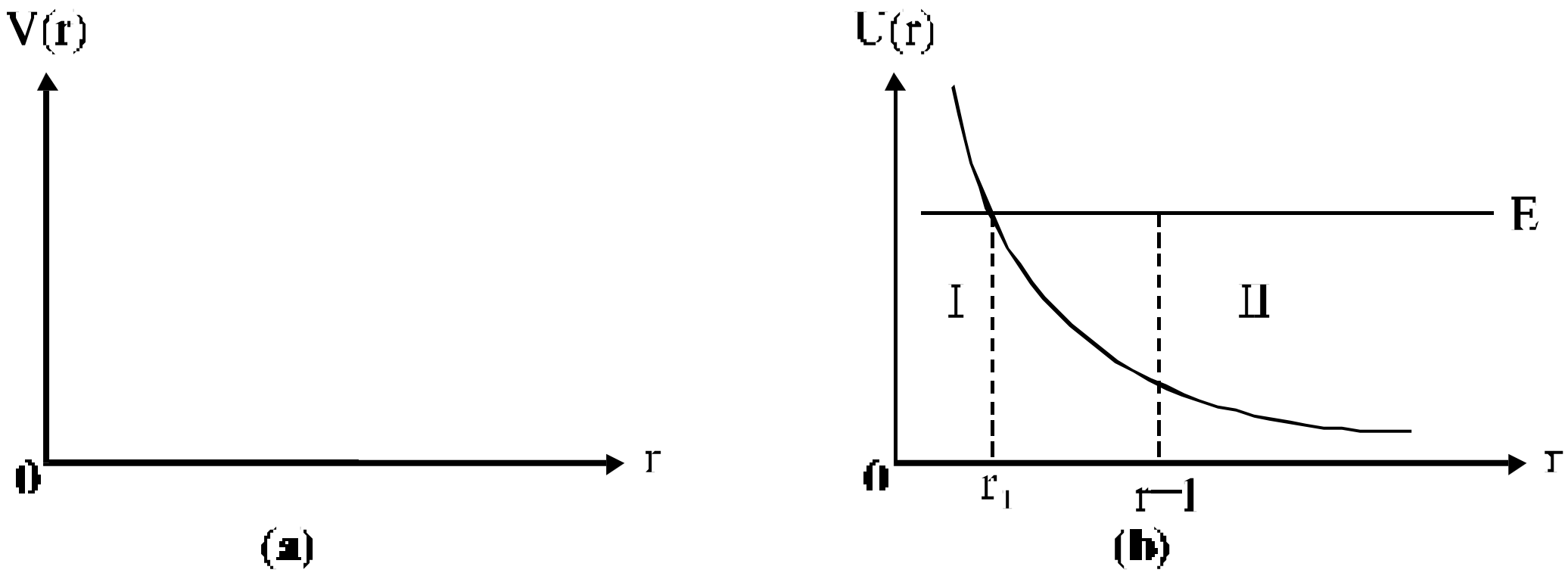

**Figure 1.** (a) The free particle potential V(r), (a = 0) (b) The free particle effective potential U(r), (a ≠ 0). Where, $r_1$ is the root of the equation E = U(r), namely, $r_1 = \sqrt{a/E}$ . (I) is the domain for $0 < r < r_1$ and (II) is the domain for $r_1 < r < +\infty$.

As seen from the Figure 1, in the domain I, the wave function should be zero, because U(r) is infinitely large. Therefore, in these domains, the wave functions, namely,

$F_1(r) = 0, \quad F_2(r) = F(r) \neq 0$

In the domain II, the energy of a free particle is positive, E > 0, and $k = im_1\sqrt{E} = iK$ , with $K = m_1\sqrt{E}$

$$G(r) = im_1\int\sqrt{U(r)}dr = im_1\int\sqrt{\frac{a}{r^2}}\,dr = im_1\sqrt{a}\ln(r)$$

$$= i\sqrt{\frac{2m}{\hbar^2}\frac{\hbar^2\ell(\ell+1)}{2m}}\ln(r) = i\sqrt{\ell(\ell+1)}\ln(r) = iQ(r), \quad \text{with} \quad Q(r) = \sqrt{\ell(\ell+1)}\ln(r)$$

The function Q(r) is depicted in Figure 2. We consider the function below

$$F(r) = f(r)\left[A_1 e^{iG(r)} + B_1 e^{-iG(r)}\right] \tag{1}$$

Where, $f(r) = e^{kr}, e^{-kr}, \ \cosh(kr), \ \sinh(kr)$.

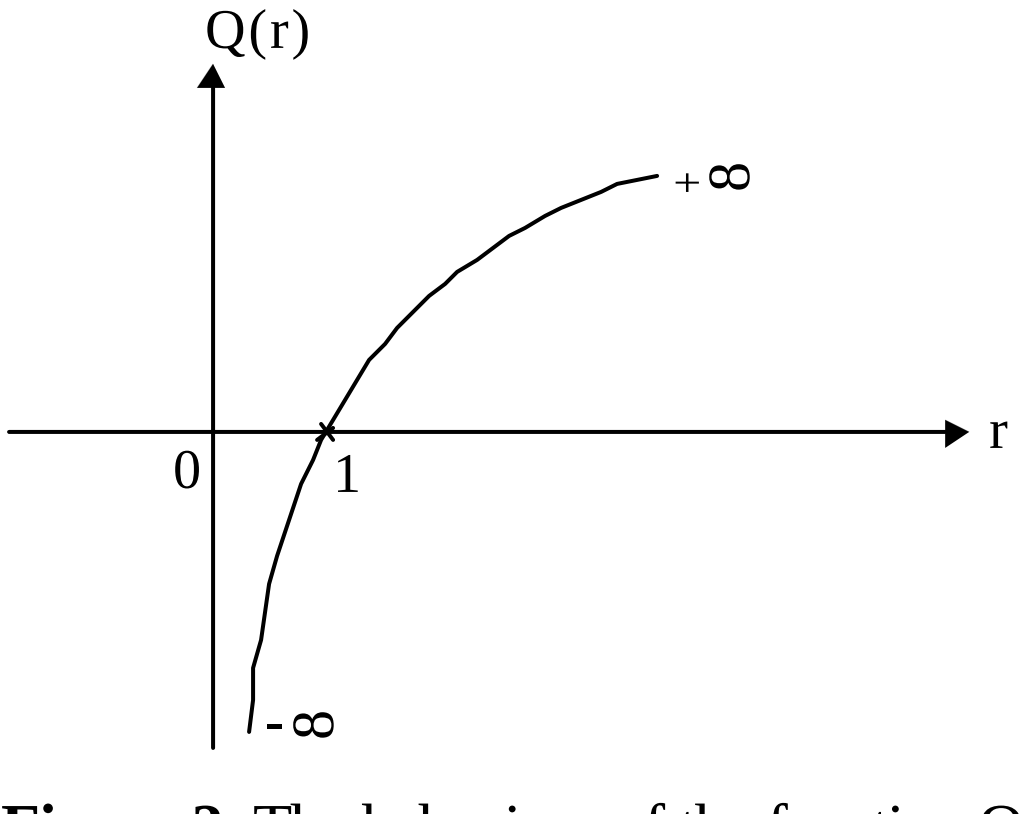

**Figure 2.** The behaviour of the function Q(r).

| | | | | |
|---|---|---|---|---|
| (1) | Q(r) = 0 | for | r = 1 | and r = 0 (because $\ell = 0$) |
| (2) | $Q(r) = -\infty$ | for | r < 1 | |
| (3) | $Q(r) = +\infty$ | for | r > 1 | |

If we put k = i K in f(r), and G(r) = i Q(r) in the equation (1). We obtain $f(r) = e^{iKr}, e^{-iKr}$, $\cos(Kr)$, $i\sin(Kr)$ and

$$F(r) = f(r)\left[A_1 e^{iG(r)} + B_1 e^{-iG(r)}\right] = f(r)\left[A_1 e^{-Q(r)} + B_1 e^{Q(r)}\right] \tag{2}$$

In order that F(r) must vanish for r = 0, $A_1$ must be zero, and in order that F(r) must vanish for $r = \infty$, $B_1$ must be zero. Thus, the wave functions are written as follows:

$$R(r) = \frac{F(r)}{r} = \frac{B_1}{r} f(r) e^{Q(r)} \qquad \text{for} \quad r < 1 \tag{3a}$$

$$R(r) = \frac{F(r)}{r} = \frac{C}{r} f(r) \qquad \text{for} \quad r = 1 \tag{3b}$$

$$R(r) = \frac{F(r)}{r} = \frac{A_1}{r} f(r) e^{-Q(r)} \qquad \text{for} \quad r > 1 \tag{3c}$$

For r = 1, these three functions have the same value. By this condition, we obtain $A_1 = B_1 = C = A$. Therefore, the wave functions are

$$R(r) = \frac{F(r)}{r} = A\frac{f(r)}{r} e^{Q(r)} \qquad \text{for} \quad r < 1 \tag{4a}$$

$$R(r) = \frac{F(r)}{r} = A\frac{f(r)}{r} \qquad \text{for} \quad r = 1 \tag{4b}$$

$$R(r) = \frac{F(r)}{r} = A\frac{f(r)}{r} e^{-Q(r)} \qquad \text{for} \quad r > 1 \tag{4c}$$

We can write the normalization condition as follows:

$$\int_0^\infty R^*(r)\cdot R(r)\, r^2\, dr = \int_0^\infty \frac{F^*(r)}{r}\cdot\frac{F(r)}{r}\, r^2\, dr = \int_0^\infty F^*(r)\cdot F(r)\, dr = 1 \tag{5}$$

r is greater than $r_1$, namely, $r_1 < r$. Thus, in the equation (5), the limit of the integral must start from $r_1$, not 0. Namely,

$$Q(r) = \sqrt{\frac{2m}{\hbar^2}} \int \sqrt{ar^2 + \frac{b}{r^2}}\, dr \tag{6}$$

From the equation (6), we can write

$$\int_{r_1}^{1} F^*(r)\cdot F(r)\, dr + \int_1^\infty F^*(r)\cdot F(r)\, dr = 1 \tag{7}$$

For example, for $f(r) = e^{iKr}$, let us try hard to calculate the coefficient A. In this case,

$$F(r) = Ae^{iKr} e^{Q(r)} \qquad \text{for} \quad r < 1 \tag{8a}$$

$$F(r) = Ae^{iKr} \qquad \text{for} \quad r = 1 \tag{8b}$$

$$F(r) = A e^{iKr} e^{-Q(r)} \qquad \text{for} \qquad r > 1 \tag{8c}$$

If we put these values of F(r) in (7), we obtain

$$AA^*\left[\int_{r_1}^{1} e^{2Q(r)}\,dr + \int_{1}^{\infty} e^{-2Q(r)}\,dr\right] = 1 \;\rightarrow\; |A|^2\left[\frac{1}{1+2a}(1-r_1^{1+2a}) + \frac{1}{2a-1} + \frac{\infty^{1-2a}}{1-2a}\right] = 1 \tag{9}$$

Where $a = \sqrt{\ell(\ell+1)}$ . From (9) we see that the coefficient A is indefinite. That is to say the wave function is not normalized.

### II.2. ENERGY VALUES

From $K = \sqrt{\frac{2m}{\hbar^2}E}$ , we have $E = \frac{\hbar^2 K^2}{2m}$. Since $E = \frac{p^2}{2m}$, we have $p = \hbar K$. Both the energy and momentum values for the free particle comprise a continuum of values.

## III. CONCLUSION

By using our method, we have solved the radial Schrödinger equation for a free particle and we have found the functions given by the equations (4). These functions are not dependent on the spherical Bessel and Neumann functions although the known functions of a free particle are dependent. We consider that our solutions are very simple functions.